\documentclass[showpacs,prl,twocolumn,preprintnumbers,amsmath,amssymb]{revtex4}
\newcommand{\beq}{\begin{equation}}
\newcommand{\eeq}{\end{equation}}

\newcommand{\id}{i\kern.06em\hbox{\raise.25ex\hbox{$/$}\kern-.60em$\partial$}}

\newcommand{\bs}{/\kern-.52em b}
\newcommand{\qs}{/\kern-.52em s}

\newcommand{\p}{\partial}

\newcommand{\dd}
{\kern.06em\hbox{\raise.25ex\hbox{$/$}\kern-.60em$\partial$}}

\newcommand{\vep}{\varepsilon}

\newcommand{\bi}{{\bf i}}
\newcommand{\bj}{{\bf j}}

\newcommand{\bk}{{\bf k}}

\newcommand{\bl}{{\bf l}}
\newcommand{\be}{{\bf e}}

\newcommand{\JA}{J_{\rm AF}}
\newcommand{\JH}{J_{\rm H}}

\newcommand{\bS}{{\bf S}}

\newcommand{\lj}{\langle}
\newcommand{\rj}{\rangle}

\input amssym.def
\input amssym
\input epsf

\usepackage{graphicx}
\usepackage{dcolumn}
\usepackage{bm}
\usepackage{epsfig}

\begin{document}
\title{Spin-wave softening and Hund's coupling in ferromagnetic manganites}
\author{Sze-Shiang Feng$^\dag$, Mogus Mochena}
\affiliation
 {Physics Department, Florida A \& M  University, Tallahassee, FL 32307}
%

\date{\today}
\begin{abstract}
Using one-orbital model of hole-doped manganites, we show with the
help of Holstein-Primakov transformation that finite Hund's
coupling is responsible for the spin-wave softening in the
ferromagnetic $B$-phase manganites. We obtain an analytical result
for the spin-wave spectrum for $\JH\gg t$. In the limit of infinte Hund's
coupling, the spectrum is the conventional nearest-neighbor
Heisenberg ferromagnetic spin-wave. The $o(t/\JH)$-order
correction is negative and thus accounts for the softening near the zone
boundary.
\end{abstract}
\maketitle

The observations of large magnetoresistence (LMR) in Nd$_{0.5}$Pb$_{0.5}$MnO$_3$ , giant magnetoresistence (GMR)
and colossal magnetoresistence(CMR) in manganites
 (${\rm R}_{1-x}{\rm A}_x{\rm MnO}_3$, R is a rare earth
element and A a divalent alkaline-earth metal) a decade ago \cite{s1}have rekindled much interest in these
materials
which have been known for half a century\cite{s2}.
Upon doping, the manganites undergo complicated transitions resulting in various magnetic,
charge-ordering and orbital-ordering phases, showing the interplay between relevant spin,
 charge and orbital degrees of freedom. In particular, magnetism and electronic transport
 are clearly correlated.
So it is widely believed that knowledge of spin dynamics
can provide important information of the underlying physics of CMR. Perring {\it et al} first measured the spin waves
in ${\rm La}_{0.7}{\rm Pb}_{0.3}{\rm MnO}_3$ for a broad range of $q$\cite{s3}. The magnon spectrum is well defined at
low temperatures and
can be accounted for by the nearest neighbor Heisenberg model.
Subsequent measurements for ${\rm Pr}_{0.63}{\rm Sr}
_{0.37}{\rm MnO}_3$ and ${\rm Nd}_{0.7}{\rm Sr}_{0.3}{\rm MnO}_3$\cite{s4}, ${\rm Nd}_{0.7}{\rm Ba}_{0.3}{\rm MnO}_3$
\cite{s5} further showed that the magnon spectrum
deviates from the Heisenberg model and becomes softened near the zone boundary. So the behavior seems a universal
phenomenon of manganites. \\
\indent As is well known, a number of interactions such as spin-orbital
coupling, Hund's coupling, antiferromagnetic coupling between core spins, Coulomb interaction and dynamic
Jahn-Teller effect
coexist in manganites. These interactions are supposed to explain the existence of different phases of doped manganites.
To explain the spin wave softening, various mechanisms were proposed.
The authors of \cite{s4} further
showed that the experimental spectrum can be reproduced reasonably well by an extended
Heisenberg model.
Furukawa \cite{s6} argued that the softening seems to be explainable by ferromagnetic Kondo lattice model with
bandwidth narrower than the Hund's coupling.
Solovyev et al\cite{s7} showed that the spin-wave behavior near the zone boundary has a purely magnetic spin
origin, and neither the lattice deformation nor the orbital ordering are required to account for the softening.
Dai {\it et al} argued that the observed magnon softening and broadening are due to
strong magnetoelastic interactions\cite{s8}.
And this magnon-phonon coupling was later treated quantitatively in\cite{s9}.
Using ferromagnetic Kondo lattice model and composite
operator method, Mancini {\it et al} obtained the softening spectrum\cite{s10}.  Shannon {\it et al} constructed a theory
of spin wave excitations in the bilayer manganite ${\rm La}_{1.2}{\rm Sr}_{1.8}{\rm Mn}_2{\rm O}_7$ based on the
simplest double exchange model and explained partly the softening behavior\cite{s11}. Krivenko {\it et al} showed that the scattering of spin excitations
by low-lying orbital modes may cause the magnon softening\cite{s12}.\\
\indent In this paper, we show that in the hole-doped manganites, the softening behavior might be of a purely
 electronic origin ,i.e., a strong but finite Hund's coupling between the $e_g$
electron and the core spin. Since in the hole-doped manganite there is less than
one $e_g$ electron per site on average and the $d_{x^2-y^2}$ orbital energy is
 significantly higher than that of $d_{3z^2-r^2}$\cite{s13} due to
Jahn-Teller splitting,
one orbital description is a reasonable approximation. As in \cite{s14}, we adopt the model Hamiltonian
\begin{widetext}
\beq
H=t\sum_{\lj\bi,\bj\rj}\sum_\sigma
c^\dag_{\bi\sigma}c_{\bj\sigma}-\JH\sum_\bi {\bf s}_\bi\cdot {\bf
S}_\bi +\JA \sum_{<\bi,\bj>}{\bf S}_\bi\cdot{\bf
S}_\bj-\mu\sum_{\bi\sigma}c^\dag_{\bi\sigma}c_{\bi\sigma}+U\sum_\bi
n_{\bi\uparrow}n_{\bi\downarrow}
\eeq
\end{widetext}
where $t$ is the double exchange hopping, $\lj\bi,\bj\rj$ are nearest sites, $\mu$ is the chemical
potential for the fermions,
 $c_{\bi\sigma}$ represents the $e_g$ electrons,
 $\JH$ is the Hund's coupling between the $e_g$ spin
${\bf s}_\bi=\frac{1}{2}c^\dag_\bi\mbox{\boldmath$\sigma$}c_\bi$ and the the core spin $\bS_\bi$. $\JA$ is the
antiferromagnetic interaction between the core spins, which is necessary to account for the $G$-phase
parent $(x=1)$ manganites. The last term
is the Hubbard Coulomb interaction. We use the Holstein-Primakoff transformation
for the core spins ($S=3/2)$;
$ S^+_\bi=(2S-a^\dag_\bi a_\bi)^{1/2}a_\bi,
S^-_\bi=a_\bi^\dag (2S-a^\dag_\bi a_\bi)^{1/2},
S^z_\bi=S-a^\dag_\bi a_\bi $
and take the approximation $(2S-a^\dag_\bi a_\bi)^{1/2}\simeq (2S-\lj a^\dag_\bi a_\bi\rj)^{1/2}$.
Homogeneity implies that $\lj a^\dag_\bi a_\bi\rj=\lj a^\dag a\rj$.   Because we consider the low temperature case, we can
drop the magnon quadratic term, hence
 the total Hamiltonian can be written
 as
\begin{widetext}
\begin{eqnarray}
H&=&t\sum_{\lj\bi,\bj\rj}\sum_\sigma
c^\dag_{\bi\sigma}c_{\bj\sigma}-\mu\sum_{\bi\sigma}c^\dag_{\bi\sigma}c_{\bi\sigma}
+U\sum_\bi n_{\bi\uparrow}n_{\bi\downarrow} -\frac{1}{2}\JH
A\sum_\bi(s^+_\bi a_\bi^\dag
+s^-_\bi a_\bi)-\JH S\sum_\bi s^z_\bi\nonumber\\
&&+A^2\JA\sum_{\lj\bi,\bj\rj}a_\bi a_\bj^\dag +\JA ZN S^2-2Z\JA
S\sum_\bi a^\dag_\bi a_\bi +\JH\sum_\bi s^z_\bi a^\dag_\bi a_\bi
\end{eqnarray}
\end{widetext}
, where $A^2=2S-\lj a^\dag a\rj,Z=6$ is the coordination number of the core spins.
To use the composite operator method, we consider the doublet
$ B(\bi)=(\begin{matrix}a_\bi, &
s^+_\bi\end{matrix})^T$. The equation of motion for $B(\bi)$ is
\begin{widetext}
 \beq
 i\p_tB(\bi)=[B(\bi),H]=\left(\begin{matrix}-\frac{1}{2}\JH
As^+_\bi+A^2\JA\sum_\be a_{\bi+\be}-2Z\JA Sa_\bi+\JH s^z_\bi
a_\bi\cr
t\sum_\be(c^\dag_{\bi\uparrow}c_{\bi+\be\downarrow}-c^\dag_{\bi+\be\uparrow}c_{\bi\downarrow})
-\JH As^z_\bi a_\bi+\JH S s^+_\bi-\JH s^+_\bi a^\dag_\bi a_\bi
\end{matrix}\right)
\eeq
\end{widetext}
Composite operator method assumes that the right-hand side can be expressed as
\beq
[B(\bi),H]=\sum_\bj \varepsilon(\bi,\bj)B(\bj)
\eeq
with $\varepsilon(\bi,\bj)$ determined in the following way,
\beq
\varepsilon(\bi,\bj)=\sum_\bl m(\bi,\bl)I^{-1}(\bl,\bj)
\eeq
where $I(\bi,\bj)=\lj [B(\bi),B^\dag(\bj)]\rj,  m(\bi,\bj)=\lj [i\p_tB(\bi),B^\dag(\bj)]\rj$,
and $\lj\rj$ represents the expectation value. Thus $\varepsilon(\bi,\bj)$ contains some
parameters to be determined self-consistently.
This approach was proposed for Hubbard model originally\cite{s14} and recent intensive
studies\cite{s15} show credible agreement with Monte Carlo method.
In our case (again due to homogeneity, $\lj s^z_\bi\rj=\lj s^z\rj$)
\begin{eqnarray}
I(\bi,\bj)&=&\delta_{\bi\bj}\cdot {\rm diag}(1,2\lj
s^z\rj)\nonumber\\
m_{11}(\bi,\bj)&=&\delta_{\bi\bj}(\JH \lj s^z\rj-2ZS\JA)+A^2\JA\sum_\be \delta_{\bj,\bi+\be}\nonumber\\
m_{12}(\bi,\bj)&=&\delta_{\bi\bj}\JH(-A\lj s^z\rj-\lj s^-_\bi a_\bi \rj)\nonumber\\
m_{22}(\bi,\bj)&=&-tp_1\sum_\be(\delta_{\bi\bj}-\delta_{\bj,\bi+\be})
+\JH A\lj s^-_\bi a_\bi\rj\nonumber\\&& +2\JH S \lj s^z\rj -2\JH\lj s^z_\bi
a^\dag_\bi a_\bi \rj\nonumber
\end{eqnarray}
where $
p_1=\sum_\sigma\lj c^\dag_{\bi\sigma}c_{\bi\sigma}\rj, p_2=\lj
s^-_\bi a_\bi \rj, p_3=\lj s^z_\bi a^\dag_\bi
a_\bi \rj $. In the $\bk$-space
\begin{eqnarray}
m_{11}(\bk)&=&(\JH\lj s^z\rj-2ZS\JA)+ZA^2\JA\gamma_\bk\nonumber\\
m_{12}(\bk)&=&\JH(-A\lj s^z\rj-p_2)\nonumber\\
m_{22}(\bk)&=&\JH Ap_2-tZp_1(1-\gamma_\bk)  +2\JH S \lj s^z\rj
-2\JH p_3\nonumber
\end{eqnarray}
We assume that at $T=0$K,$\lj a^\dag a\rj=0$, which satisfies self-consistency using the resulting retarded Green's
function and spectral theorem. Then condition $\omega_{|\bk=0}=0$ requires that
$p_3=-\frac{1}{2}p_2(A+\frac{p_2}{\lj s^z\rj})$. So the $\vep$-matrix is
\begin{eqnarray}
\vep_{11}(\bk)&=&\JH \lj s^z\rj-2ZS\JA(1-\gamma_\bk) \nonumber\\
\vep_{12}(\bk)&=&-\frac{\JH}{2\lj s^z\rj}(A\lj s^z\rj+p_2)\nonumber\\
\vep_{21}(\bk)&=&-\JH(A\lj s^z\rj+p_2) \nonumber\\
\vep_{22}(\bk)&=&\JH S-\frac{tZp_1}{2\lj s^z\rj}(1-\gamma_\bk)+\JH A\frac{p_2}{\lj s^z\rj}
            +\JH \frac{p_2^2}{2\lj s^z\rj^2}\nonumber
\end{eqnarray}
and the Green's function is
\begin{eqnarray}
D_{11}(\omega,\bk)&=&\frac{\omega-(\JH S-\frac{tZp_1}{2\lj
s^z\rj}(1-\gamma_\bk)
               +\JH A\frac{p_2}{\lj s^z\rj}+\JH \frac{p_2^2}{2\lj s^z\rj^2})
              }{(\omega-\omega_1(\bk))(\omega-\omega_2(\bk))} \nonumber\\
D_{12}(\omega,\bk)&=&D_{21}(\bk)=\frac{-\JH(A\lj s^z\rj+p_2)}{(\omega-\omega_1(\bk))(\omega-\omega_2(\bk))}\nonumber\\
D_{22}(\omega,\bk)&=&\frac{2\lj s^z\rj[\omega-(\JH \lj
s^z\rj-2ZS\JA(1-\gamma_\bk))}
{(\omega-\omega_1(\bk))(\omega-\omega_2(\bk))}\nonumber
\end{eqnarray}
where $\omega_{1,2}(\bk)$ are acoustical and optical branches of the spin excitations.
 Using
\beq
p_2=\frac{1}{N}\sum_\bk\frac{i}{2\pi}\int d\omega\lim_{\eta\rightarrow 0}\frac{D_{12}(\omega+i\eta,\bk)
-D_{12}(\omega-i\eta,\bk)}{e^{\beta\omega}-1}
\eeq
we have at $T=0$K, $p_2=0$, therefore, $p_3=0$.
Accordingly, in this scheme,there are two parameters left: $\lj s^z\rj,p_1$. The acoustical magnon spectrum
can be expanded as a Taylor series which manifests the role of Hund's coupling
\beq
\omega_1(\bk)=t\sum_{n=0}^{\infty}(\frac{t}{\JH})^{n}a_n(1-\gamma_\bk)^{n+1}
\eeq
where
as usual,$\gamma_\bk=Z^{-1}\sum_{\be} e^{i\bk\cdot\be}$.
The first few $a_n$ are
\begin{eqnarray}
a_0&=&-\frac{3(4S^2\frac{\JA}{t}+p_1)}{\lj s^z\rj +S}\nonumber\\
a_1&=&-\frac{9S(4S\frac{\JA}{t}\lj s^z\rj-p_1)^2}{\lj s^z\rj(\lj s^z\rj+S)^3}\nonumber\\
a_2&=&\frac{27(4S\frac{\JA}{t}\lj s^z\rj-p_1)^3S(S-\lj s^z\rj)}{\lj s^z\rj^2(S+\lj s^z\rj)^5}\nonumber\\
a_3&=&-\frac{81(S^2-3S\lj s^z\rj+\lj s^z\rj^2)(4S\frac{\JA}{t}\lj s^z\rj-p_1)^4S}{\lj s^z\rj^3(S+\lj s^z\rj)^7}\nonumber
\end{eqnarray}
In the small $k$ limit, $\omega\simeq Dk^2, D=-(4S^2\frac{\JA}{t}+p_1)/(\lj s^z\rj +S)$.
Note that the hopping energy $tp_1$ is negative and when it overcomes the AF term, the resulting
magnon stiffness $D$ is positive. Our numerical results show that this self-consistency is satisfied. Expression (7) suggests that the softening comes from the finite $\JH$. To fix the parameters $\lj s^z\rj$, we use
spectral theorem and get $ \lj s^z\rj=\frac{1}{2}(1-x)$, where $x$ is the dopant concentration.
To fix $p_1$, we need the fermion sector.
Using the notations in \cite{s10} for the fermion operator
$\psi(\bi)=(\begin{matrix}\xi_{\uparrow\bi},\eta_{\uparrow\bi},\xi_{\downarrow\bi},
\eta_{\downarrow\bi}\end{matrix})^T$, where $\xi_\sigma=(1-n_{-\sigma})c_\sigma,\eta_\sigma=n_{-\sigma}c_\sigma$
are the Hubbard operators, we obtain
the retarded Green's function for $\psi$ in the large$-U$ limit at zero temperature.
\beq
G^R(\omega,\bk)={\rm diag}(\frac{1}{\omega-E_1(\bk)},
0, \frac{x}{\omega-E_3(\bk)}
,\frac{1-x}{\omega-E_4(\bk)} )
\eeq
('diag' means diagonal matrix) with $E_1(\bk)=-\mu+6t\gamma_\bk-\frac{1}{2}S\JH,
E_2(\bk)=U-\mu+6tu+6tv\gamma_\bk,
E_3(\bk)=[24tx\gamma_\bk-2\mu x+S\JH
x+12tp_\downarrow\gamma_\bk-12t\gamma_\bk+12t \Delta_\uparrow ]/
(2x),E_4(\bk)=[-2Ux+2 \mu x-S\JH x+2U+12tp_\downarrow
\gamma_\bk+S\JH-2\mu+12t \Delta_\uparrow]/[2(1-x)]
$
, where $\Delta$ is related to the nearest-neighbor correlations of the Hubbard operators
: $\Delta_\sigma=\lj \xi_\sigma(\bi+\be)\xi^\dag_\sigma(\bi)\rj-\lj \eta_\sigma(\bi+\be)\eta^\dag_\sigma(\bi)\rj$.
 In this scenario, $E_1, E_3$ are partially filled
and $E_2, E_4$ are empty. The relevant  parameters are $\mu,\Delta_\uparrow,p_\downarrow$. We have three
 equations to fix them
$
1-x=2-C^F_{11}-C^F_{22}-C^F_{33}-C^F_{44},\,\,\,\,\,\,,
\Delta_\uparrow=C^{F\gamma}_{11},\,\,\,\,\,\,\,\,\,
C^F_{11}=C^F_{33}
$
, where $C^F=\lj \psi(\bi)\psi^\dag(\bi)\rj, C^{F\gamma}=\lj \psi(\bi+\be)\psi^\dag(\bi)\rj$. We
 know that $C^F_{22}=0 $ and $
C^F_{44}=1-x$. Thus $C^F_{11}=x=C^F_{33}$, so $E_3$ is empty,
i.e., only $E_1$ is partially filled.  Hence only $\mu$ is
relevant to our problem and can be fixed by
$ x=N^{-1}\sum_\bk
\theta(E_1(\bk))
$ where $\theta(x)$ is the usual step function.
 The hopping energy is
\beq
tp_1=-tC^{F\gamma}_{11}=\frac{t}{N}\sum_\bk
\theta(-E_1(\bk))\gamma_\bk<0
\eeq
The other two parameters $\Delta_\uparrow, p_\downarrow$ can also be determined by
$t\Delta_\uparrow=-tp_1>0\nonumber,
24tx+12t(p_\downarrow-1)=-2\mu x+S\JH x+12t\Delta_\uparrow
$
. Further analysis  show that for $\JH>2.5t$, the whole scheme is
self-consistent. Fig.1 shows that two relevant fermion bands for
$x=0.301,t=1,\JH=3.0$ ( in unit of $t$).\\
\indent It is seen from the magnon spectrum (7) that we can estimate the two model
parameters $t$ and $\JH$ from measured data. Fig.2 shows the
comparison between our calculated result for the prescribed antiferromagnetic coupling $\JA=0.01$
and the measured result at $T=10$K for ${\rm Pr}_{0.63}{\rm Sr}
_{0.37}{\rm MnO}_3$ in\cite{s4}. The solid curve in the left panel is the fit to a nearest
-neighbor Heisenberg model and gives the value at zone boundary about 34.2meV. This corresponds to the the uppermost
curve in the right panel. The comparison gives the hopping energy $t\simeq 0.462$eV.  The circles are the
data measured and give the value at zone boundary about 23meV, corresponding to the point 0.05 in the right panel.
This point
corresponds to  $\JH\simeq=3.2t\simeq 1.48$ eV.
Note that the ratio $\JH/t$ is very close to the values of interaction from a number of references\cite{s16}\cite{s17}.
It is worth noting that the nearest-neighbor Heisenberg interaction alone can not account for
the Curie temperature. The fitting curve in the left panel corresponds to the nearest-neighbor Heisenberg
spectrum $\omega(\bk)\simeq 51.3(1-\gamma_\bk)$ meV. In the mean field theory,the Curie temperature $T_c$
corresponding to the spectrum $\omega_\bk=2ZS^*J(1-\gamma_\bk)$ is $k_BT_c=\frac{2}{3}JZS^*(S^*+1)$ (here
$S^*=S+\frac{1}{2}(1-x)$ is the effective spin). This gives $T_C^{\rm MF}\simeq 500$K. Taking into account that in three dimensions for a simple cubic lattice
, the real Curie temperature $T_C$ and $T_C^{\rm MF}$ have relation\cite{s18}: $T_C=0.75 T^{\rm MF}_C$, we get
$T_c\simeq 375$ K, higher than the real value 315 K.
\begin{figure}
\epsfxsize=8truecm {\epsfbox{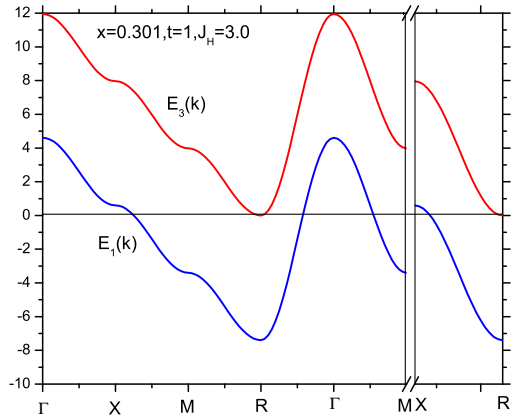}}
\caption{\label{fig:epsart}The bands for the Hubbard operators at $T=0$K.}
\end{figure}
\begin{figure}
\begin{minipage}[h]{3.cm}
\vskip -1.mm
\hskip -6cm
\epsfxsize=3.15truecm {\epsfbox{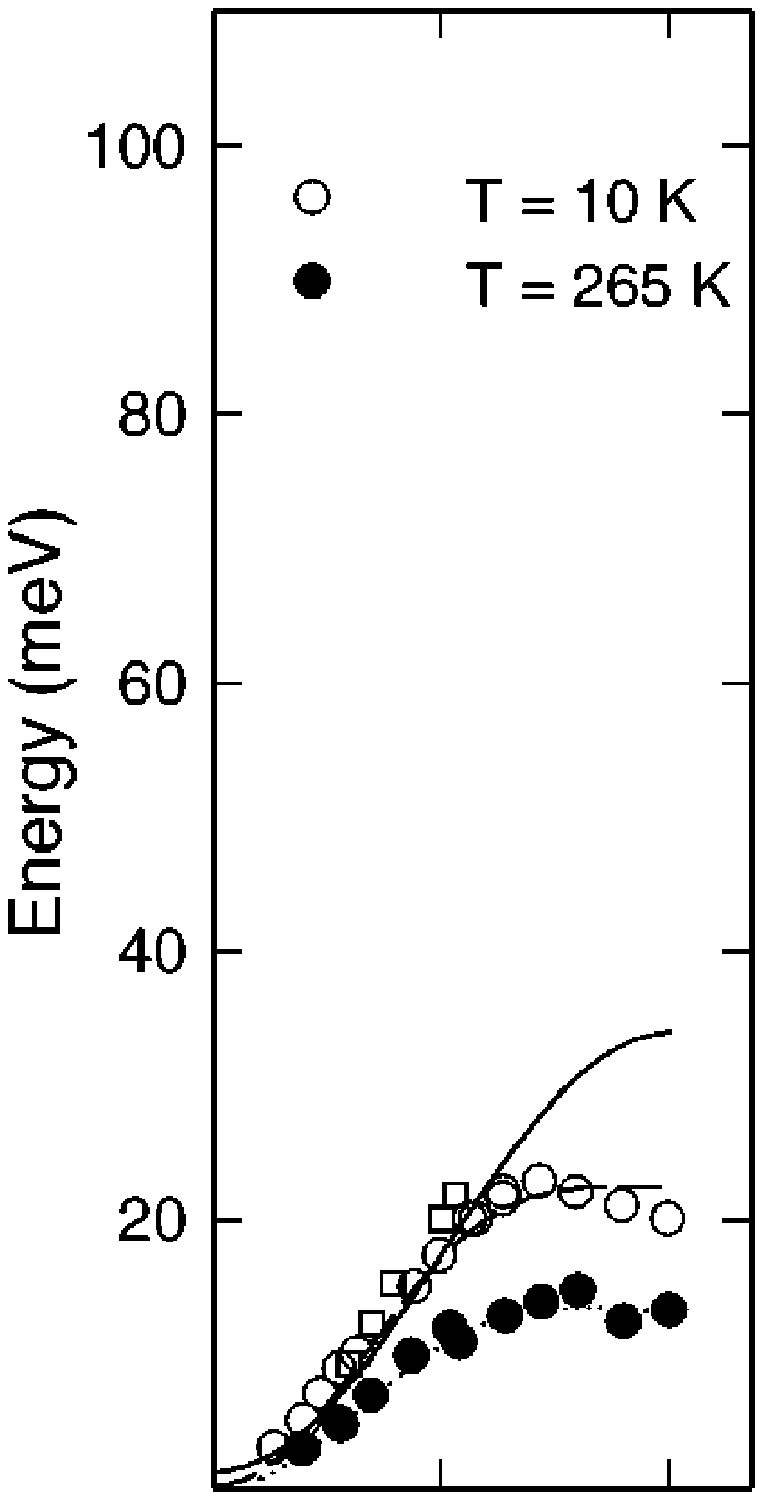}}
\end{minipage}
\hskip -3.8cm
\begin{minipage}[h]{3.5cm}
\epsfxsize= 6.8truecm
 {\epsfbox{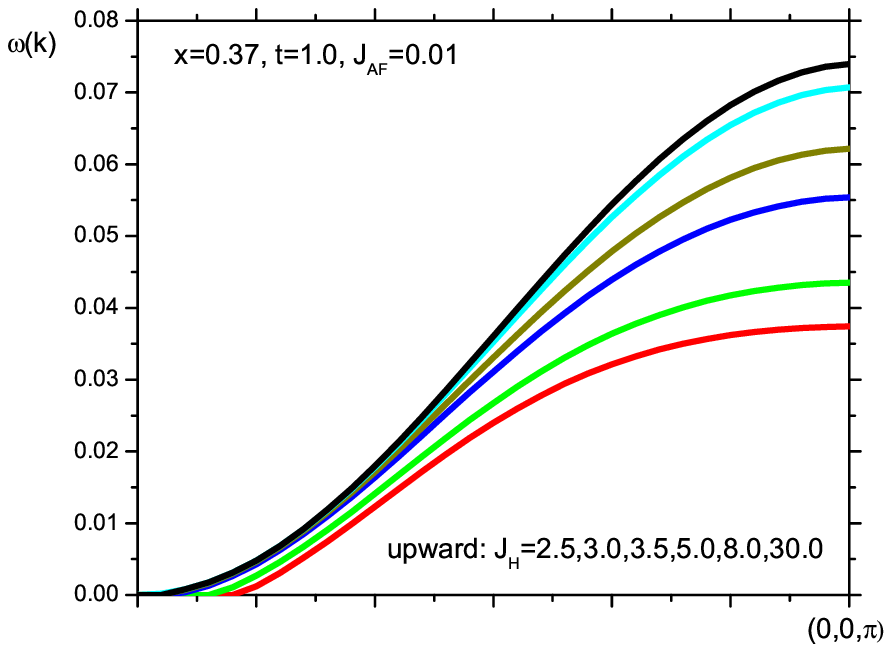}}
\end{minipage}
\caption{The spin wave spectrum along $\Gamma X$ of $B$-phase. The left panel (taken from\cite{s4})
 is the experimental result.The right panel is the
calculated spin-wave spectrum (in unit of $t$) at $T=0$K .}
\end{figure}
 \indent To conclude this paper, we present some discussions and comments.
In the derivation of the series expression of the magnon spectrum,
  we have used the approximation
$(2S-a^\dag_\bi a_\bi)^{1/2}\simeq (2S-\lj a^\dag_\bi a_\bi\rj)^{1/2}$
in the Holstein-Primarkov transformation. This can be satisfied at very low temperatures. Further, the quartic term
$\JA\sum_{\lj\bi,\bj\rj} a^\dag_\bi a_\bi
a^\dag_\bj a_\bj$  is neglected because  $\JA$ is very small and the magnon fluctuation at
zero temperature is negligible. The series  expression (7) of the accoustic magon dispersion shows
 alternating behavior;  convergence is guaranteed when
$\JH S>3$.
The model parameters $t,\JH,\JA$  and the hopping energy $p_1$ can be estimated by fitting
experimental data.
  There is a simple physical picture for the deviation of the magnon spectrum from that of Heisenberg
  model.  The interaction between core spins is induced by the hopping of $e_g$-electrons and the dominant term
  is linear in $t$. If
 the Hund's coupling $\JH$ is infinite,only the dominant term plays the role.
 The $e_g$ electron and core spin must add up to a total spin-2 to minimize the energy
 in the $B$-phase.  So the actual background
 for spin excitation is just that in the simple Heisenberg model. But for finite $\JH$, high orders
 of the mediated interaction between core spins make some difference.
Our result (7) agrees with the conclusions from random phase approximation\cite{s19}, which provides an integral
 equation for the dispersion relation. The strength of induced
 ferromagnetic interaction is determined by the hopping energy of the conduction fermions. For the approach presented
 to be self-consistent, the spin-wave stiffness must be positive. The ferromagnetic order becomes unstable
 at certain filling when the the stiffness vanishes.
  Though zone boundary spin wave softening can be explained by the spin dynamics in the
  Kondo lattice model, as shown in this paper. The origin of the behavior is still an issue of debate.
  Based on the observed proximity of phonon dispersion and magnon dispersion and the anisotropic spin-wave
  broadening, Dai et al\cite{s8}
  argue that strong magnon-phonon coupling is needed for a complete understanding of the
  low temperature spin dynamics of manganites.
Quite recently, Endoh et al concluded\cite{s20} that the
ferromagnetic magnons in Sm$_{0.55}$Sr$_{0.45}$MnO$_3$  is of orbital nature since the magnon dispersion
shows anisotropy which is mainly determined by the short range correlation of the $e_g$ orbitals. They explained
that the anisotropic magnon dispersion are attributed to long range magnetic
interactions based on fitting the data to Heisenberg model with long range interactions.  We believe that
if orbital degrees of freedom are taken into account in our model, the resulting magnon spectrum will be anisotropic
since orbital degrees of freedom bring anisotropy into the system.
Finally, we remark that as manganites are very complex systems, there might be multiple mechanisms contributing to
a single phenomenon.  The analysis provided in this paper shows that Hund coupling might be of primary importance.
 \\
\indent SSF is grateful to Prof. P. Schlottmann, Prof. Mancini, and Dr. Avella for their
helpful discussions. This work is supported in part by the Army High Performance Computing Research
Center(AHPCRC) under the auspices of the Department of the Army, Army Research Laboratory (ARL) under
Cooperative Agreement number DAAD19-01-2-0014. \\
$^\dag$Corresponding author,e-mail:shixiang.feng@famu.edu

\end{document}